\documentclass[
twocolumn,
pre,
a4paper,
nofootinbib,
amsmath,
amssymb,
superscriptaddress,
showkeys,
floatfix,
notitlepage,
longbibliography
]{revtex4-2}

\usepackage[english]{babel}
\makeatletter\AtBeginDocument{\let\@elt\relax}\makeatother
\usepackage{graphicx, amsfonts, amssymb,amsmath}
\usepackage{hyperref}
\usepackage{colortbl}
\usepackage{float}
\usepackage{ulem}
\usepackage{commath}
\usepackage[sort&compress,capitalize]{cleveref}
\usepackage{siunitx}

\def\papertitle{Optimising nanoporous supercapacitors for heat-to-electricity conversion}

\usepackage{grffile}
\usepackage{siunitx}
\usepackage{xspace}

\usepackage[draft,inline,nomargin]{fixme}
\fxsetup{theme=color,mode=multiuser}
\FXRegisterAuthor{sk}{ask}{\color{blue}SK}
\FXRegisterAuthor{mj}{amj}{\color{red}MJ}
\FXRegisterAuthor{tv}{atv}{\color{red}TV}
\FXRegisterAuthor{ak}{aak}{\color{red}AK}

\usepackage{amssymb}
\usepackage{amsmath}
\usepackage{commath}
\usepackage{graphicx,bm}
\usepackage[dvipsnames]{xcolor}

\usepackage[utf8]{inputenc}

\newcommand{\latin}[1]{{\it #1}}
\newcommand{\ie}{\latin{i.e.}\@\xspace}

\newcommand{\cf}{\latin{cf.}\@\xspace}
\newcommand{\etc}{\latin{etc.}\@\xspace}

\newcommand{\vs}{\latin{vs}.\@\xspace}

\newcommand{\rhomax}{\rho_\mathrm{max}}

\newcommand{\ftitle}[1]{{\bf #1}}
\newcommand{\fsub}[1]{({\bf #1})}

\newcommand{\ESIname}{Supplementary Material\@\xspace}

\newcommand{\sfig}[1]{\cref{sm:#1} in the \ESIname}

\newcommand{\ssect}[1]{Section~\ref{sm:#1} in the \ESIname}

\newcommand{\mycite}[1]{Ref.~\cite{#1}}

\newcommand{\cit}[1]{Ref.~\cite{#1}}

\newcommand{\e}{\mathrm{e}}
\newcommand{\nn}{\nonumber\\}

\DeclareMathAlphabet{\mathbb}{U}{bbold}{m}{n}
\newcommand{\porer}{R} 
\newcommand{\poreri}{R_\delta} 
\newcommand{\porera}{R_a} 

\newcommand{\ionr}{a}  
\newcommand{\carbon}{\mathrm{c}}  
\newcommand{\carbonr}{a_\carbon}  
\newcommand{\rhoc}{\rho_\carbon}  

\newcommand{\uu}{u}  
\newcommand{\image}{\mathrm{image}}
\newcommand{\vdW}{\mathrm{vdW}}

\newcommand{\kBT}{k_{B}T}

\newcommand{\lB}{\lambda_B}

\newcommand{\epsLJ}{\epsilon_\mathrm{LJ}}

\newcommand{\oneD}{\mathrm{1D}}

\newcommand{\muoneD}{\mu^\oneD}

\newcommand{\cold}{\mathrm{L}}
\newcommand{\hot}{\mathrm{H}}

\graphicspath{ {./figures/} }

\usepackage{xr}
\begin{document}

\title{\papertitle}

\let\thefootnote\relax\footnotetext{This paper is dedicated to the memory of Professor Douglas Henderson.}

\date{\today}

\author{Mathijs Janssen}
\affiliation{Department of Mathematics, Mechanics Division, University of Oslo, N-0851 Oslo, Norway}

\author{Taras Verkholyak}
\affiliation{Institute for Condensed Matter Physics, NASU, Ukraine}

\author{Andrij Kuzmak}
\affiliation{Department for Theoretical Physics, I. Franko National University o
f Lviv, Lviv, Ukraine}

\author{Svyatoslav Kondrat}
\email{svyatoslav.kondrat@gmail.com}
\email{skondrat@ichf.edu.pl}
\affiliation{Institute of Physical Chemistry, Polish Academy of Sciences, 01-224 Warsaw, Poland}

\affiliation{Institute for Computational Physics, University of Stuttgart, Stuttgart, Germany}

\begin{abstract}

Innovative ways of harnessing sustainable energy are needed to meet the world's ever-increasing energy demands.
Supercapacitors may contribute, as they can convert waste heat to electricity through cyclic charging and discharging at different temperatures.  
Herein, we use an analytically-solvable model of a cylindrical pore filled with a single file of ions to identify optimal conditions for heat-to-electricity conversion with supercapacitors.
We consider Stirling and Ericsson-like charging cycles and show that the former or latter yields more work when a supercapacitor operates under charge or voltage limitations, respectively.
Both cycles yield the most work for pores almost as narrow as the size of the ions they contain, as is the case for energy storage with supercapacitors. 
In contrast to energy storage, which can be maximised by ionophobic pores, such pores do not yield the best heat-to-electricity conversion, independently of the applied potential. 
Instead, we find that for a given pore size, a moderately ionophilic pore harvests more work than ionophobic and strongly ionophilic pores.

\end{abstract}

\maketitle

\section{Introduction}

Undesired heat is ubiquitous in industry, data centers, and electronic devices such as smartphones and laptops. 
Different methods have been proposed recently to harness this energy, many of which rely on electrochemistry, for instance, ion fluxes in thermal gradients or the temperature-dependent charging of batteries, supercapacitors, and other electrochemical devices \cite{lee2014electrochemical,kim2016thermally,gao2018engineering,brogioli2021innovative,zhao2021ionic}.

Supercapacitors store energy through the voltage-induced adsorption of ions into the pores of two, usually nanoporous, electrolyte-immersed electrodes (\cref{fig:model}a) \cite{miller08a, simon08a, beguin14a, gonzalez16a,park2022insitu}. 
Besides energy storage, supercapacitor technology is utilised for capacitive deionisation of saline water \cite{porada13a,suss18a, zhang20a} and blue energy harvesting \cite{janssen2014boosting, ahualli2014temperature,zhang2017combined, atlas2018periodic,kim2022continuous}. 
\citeauthor{hartel2015heat} \cite{hartel2015heat} showed with experiments and classical density functional theory computations that the open-circuit voltage of a supercapacitor increases with temperature, which is the essential ingredient for heat-to-electricity conversion (HEC) with supercapacitors.
(This thermal voltage rise can also boost the work output of blue-energy devices~\cite{janssen2014boosting, ahualli2014temperature,zhang2017combined, atlas2018periodic,kim2022continuous}.)
Later studies of HEC with supercapacitors included theoretical works to improve power output and efficiency \cite{wang2017thermal,lin2019performance} as well as experimental work improving the design of the HEC apparatus \cite{wang2017maximal}. 
In an extensive recent work, \citeauthor{kim2022study} compared HEC charging cycles of a commercial pseudocapacitor, Li-ion supercapacitor, and an electric double-layer capacitor \cite{kim2022study}.
All three types of devices had pros and cons in terms of their work output, energy loss, efficiency, operating temperature window, \etc
Still, none of these commercial storage devices were optimized with HEC in mind.

For HEC with supercapacitors to be upscaled and applied, it is imperative to find optimal electrolytes and electrodes, though the parameter space spanned by the possible combinations is vast.
Challenging in experiments, this search can be aided by theoretical models. 
Many supercapacitors models, including those by Douglas Henderson and his collaborators \cite{Busath2004, Kong2015, Schmickler2017Charge, Schmickler2017capacitance, SilvestreAlcantara2015}, focused on the charging of narrow cylindrical pores \cite{Verkholyak2021, kornyshev:fd:14, lee:nanotech:14, lee:prl:14, rochester:jpcc:16, Schmickler2015simple, pak_hwang:jpcc:16:IonTrapping, Busath2004, SilvestreAlcantara2015, Kong2015, Schmickler2017capacitance, Schmickler2017Charge, qiao2018modeling, Kondrat2019a, ZaboronskyKornyshev2020Ising}.
The behaviour of ions in such pores can be captured by one-dimensional models that have analytically-tractable solutions \cite{kornyshev:fd:14, lee:prl:14, Schmickler2015simple, rochester:jpcc:16, Verkholyak2021}.
In an earlier work, we developed an exactly-solvable off-lattice model for the charging of single-file pores \cite{Verkholyak2021}. 
Our model is computationally cheap and reproduces three-dimensional Monte Carlo simulations over a wide range of parameters. 
We use this single-file pore model here to search, for the first time, the electrode-electrolyte parameter space for combinations optimal for HEC. 
We focus on two parameters: the pore radius and the strength of ion-electrode dispersion interactions.
Taken together, these parameters affect the pore's ionophilicity---the tendency of pores to be filled in absence of an applied potential \cite{kondrat:nh:16, lian_wu:jpcm:16:Ionophobic, vos2022electric}---and in this way relate HEC optimality to ionophilicity.

This paper is organised as follows. 
We first revisit the thermodynamic principles of HEC in \cref{sec:general} and introduce our supercapacitor model in \cref{sec:model}. 
In \cref{sec:results}, we discuss for which pore radii and dispersion interactions HEC with this model supercapacitor is optimal and which thermodynamic cycle is preferred under which conditions.
We summarise and conclude in \cref{sec:concl}.

\section{Thermodynamics of HEC}
\label{sec:general}

\begin{figure}
\includegraphics[width=\linewidth]{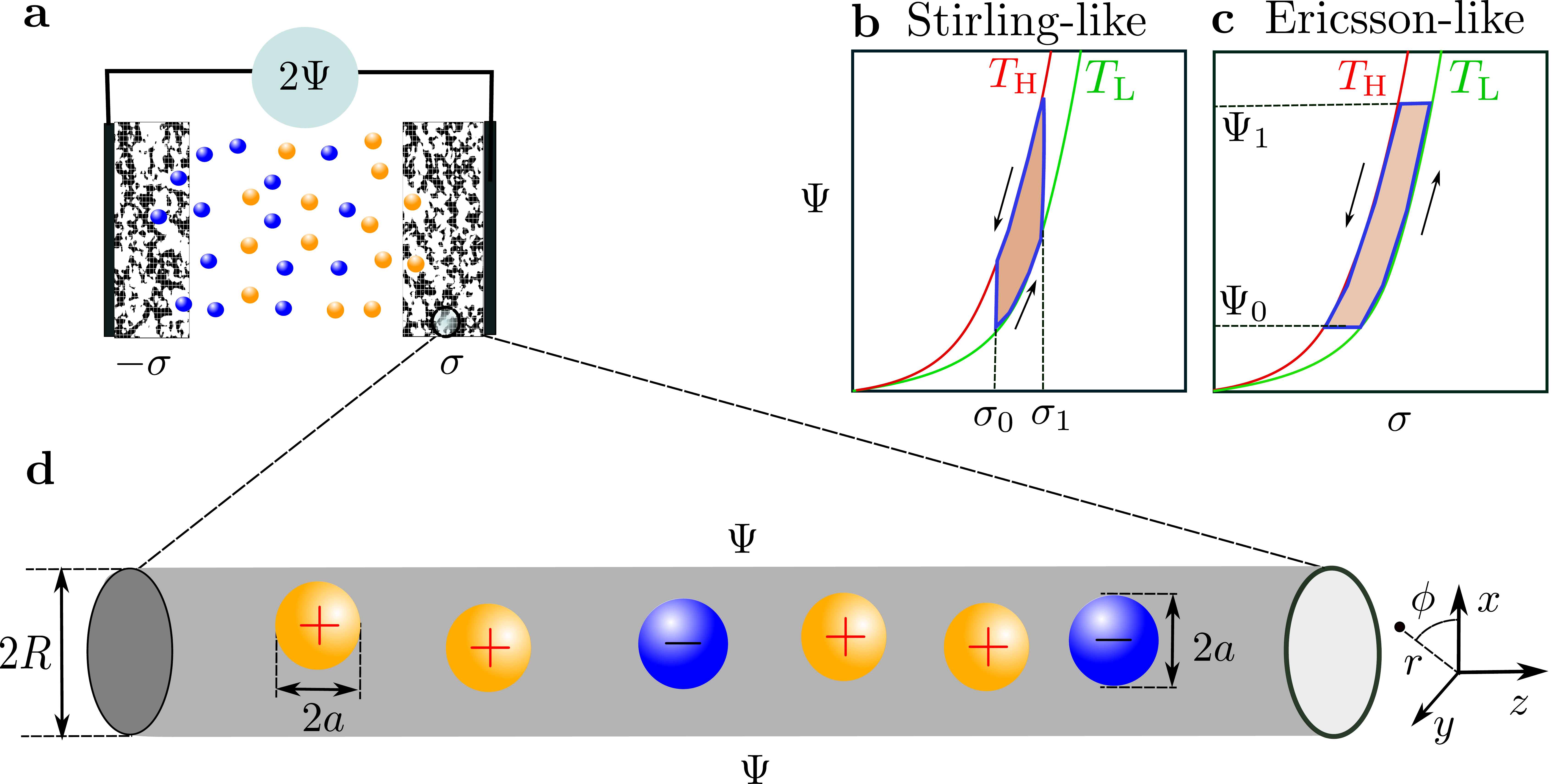}
\caption{\ftitle{Nanoporous supercapacitor for HEC.} 
	\fsub{a} Schematic of a supercapacitor consisting of two electrodes immersed in an electrolyte. 
	An electrostatic potential difference $2\Psi$ is applied between the electrodes, inducing surface charge densities $\sigma$ and $-\sigma$ on the two electrodes.
	\fsub{b,c} Schematics of HEC with Stirling-like \fsub{b} and Ericsson-like \fsub{c} charging cycles. 
	The shaded areas show the work done through the cycles and the arrows show the cycling directions.
	\fsub{d} Schematic of a cylindrical nanopore of radius $\porer$ within a supercapacitor electrode. 
	Cations and anions have the same radius $\ionr$.
	\label{fig:model}
	}
\end{figure}

We consider a supercapacitor comprising two identical porous blocking electrodes filled with an electrolyte at temperature $T$ (\cref{fig:model}a).
A potentiostat imposes an electrostatic potential difference $2\Psi$ between the electrodes, charging them to surface charges $Q$ and $-Q$. 
From hereon, we scale all extensive quantities to the surface area $A$ of one electrode; in particular, we use the surface charge density $\sigma=Q/A$.
The First Law of thermodynamics for the supercapacitor reads $dU=T\dif S-\delta W$, with $S$ being its entropy and $\delta W=-\Psi \dif \sigma$ the work done by the capacitor on its surroundings \cite{boon2011blue}.
By performing a charging-discharging cycle during which temperature is varied, the supercapacitor harvests an amount of electrical work given by \cite{janssen2014boosting, hartel2015heat}
\begin{align}
    \label{eq:work}
	W=-\oint \Psi(\sigma,T)\dif \sigma\,.
\end{align}
Comparing the First Law of the supercapacitor to that of a ``classical" heat engine---$dU=TdS-pdV$, with $p$ and $V$ being the pressure and volume of a gas-filled cylinder---we note that the same roles are played by $-p$ and $\Psi$ and by $V$ and $\sigma$. 
Charging-discharging cycles of supercapacitors are thus analogous to the contraction-expansion cycles of heat engines.
A heat engine exploits the temperature dependence in the equation of state $p(V,T)$ to convert heat to mechanical work;
a supercapacitor exploits the temperature dependence in the relation $\Psi(\sigma,T)$ to convert heat into electrical work.
While heat engines perform clockwise cycles in $p-V$ diagrams to perform work, due to the minus sign in \cref{eq:work}, HEC with supercapacitors requires counterclockwise cycles in $\Psi-\sigma$ diagrams.

\subsection{Seebeck coefficient}

In which order one should charge, discharge, heat, and cool to realise a counterclockwise cycle in a $\Psi-\sigma$ diagram depends on the sign of the electrolytic Seebeck coefficient, 
\begin{align}
	\label{eq:Seebeck}
	\mathcal{S}(\sigma,T)=\frac{\partial\Psi(\sigma,T)}{\partial T}.
\end{align}
Note here that $\Psi(\sigma,T)$ depends parametrically on the electrode's dimensions and morphology and the ion concentration, composition, relative permittivity, \etc---the next section presents our microscopic model for $\Psi(\sigma,T)$.
Prior modelling studies on the response of the electric double layer to varying temperature used (ion size-modified) mean field theories \cite{Janssen2014,ju2017analysis,chen2018temperature}, lattice models \cite{Kornyshev2013}, density functional theory \cite{reszko2005temperature,hartel2015heat}, and molecular simulations \cite{boda1999low,boda2000capacitance,chen2018temperature}.
Positive Seebeck coefficients were often found, so we will refer to $\mathcal{S}>0$ as ``normal'' and $\mathcal{S}<0$ as ``anomalous''.
Yet, in \ssect{sec:GC} we show that, even within the Gouy-Chapman model, both $\mathcal{S}>0$ and $\mathcal{S}<0$ are possible. 
We find there that $\mathcal{S}>0$ if the electrolyte's relative permittivity $\varepsilon_r$ satisfies $\partial \varepsilon_r/\partial T<\varepsilon_r/T$, which is the case for most electrolytes \cite{ju2017analysis}. 
Meanwhile, experiments on pseudocapacitors \cite{kim2022study} and electric double-layer capacitors \cite{hartel2015heat,wang2017thermal,kim2022study} yielded positive Seebeck coefficients of the order of $\mathcal{S}=\SI{0.5}{\milli\volt\per\kelvin}$.
Conversely, negative Seebeck coefficients were found for batteries \cite{lee2014electrochemical,gao2018engineering} and supercapacitors  based on Li-ion intercalation \cite{kim2022study}.
The experimentally-determined $\mathcal{S}$ varied substantially with the applied potential (Fig. (9) in \cite{kim2022study}) but hardly with temperature (Fig. S5 in \cite{hartel2015heat}). 
With a slight abuse of terminology, therefore, we will speak in the following of positive and negative Seebeck coefficients even if we compare $\Psi(\sigma,T)$ at two widely separated temperatures.

\subsection{Stirling and Ericsson-like cycles}

We consider $\mathcal{S}>0$ in the remainder of this section, for which charging-heating-discharging-cooling cycles yield positive work.
The well-known theoretical cycles of classical heat engines are defined in terms of the quantities held fixed at the different stages, be it volume (isochoric), pressure (isobaric), temperature (isothermal), or entropy (isentropic).
Analogously, we consider a Stirling-like charging cycle where temperature is changed at fixed charge (``isogalvanic") and a Ericsson-like cycle wherein temperature is changed at fixed potential (``isovoltaic").
In principle, a Carnot-like cycle with isentropic temperature changes is possible as well \cite{schiffer2006heat,janssen2017coulometry,janssen2017reversible}. 
However, the heating-by-charging and cooling-by-discharging effect that such a cycle would employ is probably too small for practical applications. 

The Stirling-like cycle operates between low and high temperatures $T_\cold$ and $T_\hot$ and between low and high charges $\sigma_0$ and $\sigma_1$ (\cref{fig:model}b).
In this case, \cref{eq:work} amounts to
\begin{align}
\label{eq:work:sigma}
W_\sigma =
	\int_{\sigma_{0}}^{\sigma_{1}}\Psi(\sigma,T_\hot)\dif \sigma 
	-\int_{\sigma_{0}}^{\sigma_{1}}\Psi(\sigma,T_\cold)\dif \sigma\,.
\end{align}
Note that these integrals represent the energies stored in a supercapacitor when it is charged from the surface charge density $\sigma_0$ to $\sigma_1$ at temperatures $T_\hot$ and $T_\cold$; the harvested energy is the difference between these energies.

The Ericsson-like cycle operates between low and high temperatures $T_\cold$ and $T_\hot$ and between low and high potentials $\Psi_0$ and $\Psi_1$ (\cref{fig:model}c). 
In this case, \cref{eq:work} amounts to
\begin{align}
\label{eq:work:psi}
W_\Psi &= 
	\int_{\Psi_0}^{\Psi_1} C(\Psi, T_\hot) \Psi \dif\Psi
	+ \Psi_1 \Delta \sigma (\Psi_1)\nn
&\quad	- \int_{\Psi_0}^{\Psi_1} C(\Psi, T_\cold)\Psi \dif\Psi
	- \Psi_0 \Delta \sigma (\Psi_0) ,
\end{align}
where $C(\Psi, T)$ is the differential capacitance and $\Delta \sigma (\Psi)= \sigma (\Psi, T_\cold) - \sigma (\Psi, T_\hot)$ is the change in the accumulated charge when heating up. 
The integral $ \int_{\Psi_0}^{\Psi_1} C(\Psi, T) \Psi \dif\Psi$ is the energy stored in an electrode by charging it from the potential difference $\Psi_0$ to $\Psi_1$ at temperature $T$, and $\Psi \Delta \sigma$ is the energy gained or lost by the system during isovoltaic heating or cooling.

Typically, supercapacitors have to operate within prescribed temperature and voltage windows.
The schematics in \cref{fig:model}b and c suggest that, given the same admissible high and low potentials, an Ericsson-like cycle yields higher work output than a Stirling-like cycle.
If, conversely, a supercapacitor's surface charge density should remain within certain bounds, a Stirling-like cycle is preferable instead.
\label{ref:2:efficiency}The efficiencies of both types of cycle are below the theoretical limit $\eta=1-T_\hot/T_\cold$ \cite{hartel2015heat}\footnote{Expressions for their efficiency involve the supercapacitor's heat capacity, which depends on its charging state \cite{park2022insitu}.}.
To approach that limit, one should either capture the heat discarded by the engine when it cools to and charges at $T_\cold$ and reuse it during the heating stage, or one should use the mentioned isentropic charging and discharging to change temperatures.
One can explicitly show that, irrespective of the ``equation of state" $\Psi(\sigma,T)$, a Carnot-like charging cycle reaches Carnot efficiency (Sec. 6B of \cite{janssen2017electric}).

\section{Microscopic model}
\label{sec:model}

To study temperature effects and the possibility of energy harvesting with supercapacitors, we consider a charged hard-spheres model for the electrolyte and a metallic nanotube as part of a supercapacitor electrode. 
We assume that the nanotube is so narrow that it can only accommodate a single file of ions. 
\citeauthor{Verkholyak2021}\cite{Verkholyak2021} mapped this model onto a one-dimensional model with an exact analytical solution for the voltage-dependent in-pore ion densities, showing a surprisingly good agreement with Monte Carlo simulations of the full (three-dimensional) model. 
Thus, the 1D model is a convenient tool to systematically study various aspects of supercapacitor charging. 
We first describe the details of the full model and then discuss the results of the mapping.

\subsection{Model}
We consider a cylindrical, metallic nanopore of radius $\porer$, extending infinitely along the axial direction (\cref{fig:model}d). 
We apply a potential difference $\Psi$ between the pore wall and a bulk electrolyte with which the pore is in contact (\cref{fig:model}a). 
The nanopore is filled with ions, which we model as monovalent charged hard spheres of equal radius $\ionr$.

\subsubsection{Ion-ion interactions inside a metallic nanotube}
\label{sec:model:Uij}

The electrostatic interaction energy between two ions in a cylindrical metallic nanopore can be expressed exactly as \cite{rochester:cpc:13}
\begin{align}
    \label{eq:Uel:axis}
	\beta \uu^{\rm el}_{\alpha\gamma} (z) =  \frac{2\lB Z_\alpha Z_\gamma}{\poreri} 
        \sum_{n=1}^\infty \frac{\e^{-k_{n0} z/\poreri}}{k_{n0}[J_{1}(k_{n0})]^2},
\end{align}
where $z$ is the distance between the ions, 
$J_{m}(x)$ are Bessel functions of the first kind of order $m$, $k_{n0}$ are zeros of $J_0(x)$,
and where, for our purposes (see \cref{sec:model:1D}), we restricted the ions to lie on the symmetry axis of the cylinder. 
Here, $Z_\alpha$ are the ionic valencies ($=\pm1$ for monovalent ions considered here), $\lB=\beta e^2/(4\pi\varepsilon_0\varepsilon_r)$ is the Bjerrum length, $\varepsilon_0$ is the vacuum permittivity, $e$ the proton charge, $\beta=1/\kBT$ is the inverse thermal energy, with $k_B$ being Boltzmann's constant, and $\poreri=\porer - \delta$ is the location of the image surface. 
We took $\delta = \SI{0.8}{\angstrom}$ \cite{Cagle2009, Lang1971}, which is slightly less than half of the radius of the wall (carbon) atom, $\carbonr = \SI{1.685}{\angstrom}$. 
\label{ref:3:screening}Note that the exponential screening of the bare Coulomb interactions between ions in \cref{eq:Uel:axis} is through the actions of electrons (or holes) on the nanotube, rather than through other ions, as in the case of a Debye screening cloud in bulk electrolytes.
This screening makes it possible to stack many same-signed ions in a narrow pore, leading to the so-called superionic state \cite{kondrat:jpcm:11}.

For large ion-ion separations, $z \gg \poreri$, \cref{eq:Uel:axis} reduces to its leading-order term,
\begin{align}
    \label{eq:Uel:approx}
	\beta \uu^{\rm el}_{\alpha\gamma} (z) \approx  \frac{3.08\lB Z_\alpha Z_\gamma}{\poreri} \e^{-2.4 z/\poreri}.
\end{align}
The above expression describes the ion-ion interactions remarkably well even when ions are close to contact ($z= 2a$) \cite{kornyshev:fd:14}, so we have used it in all calculations discussed below.


In most calculations, we used a temperature-independent relative permittivity $\varepsilon_r=35$, which corresponds to acetonitrile at room temperature; however, we also tested some of our results with temperature-dependent permittivity $\varepsilon_r (T) = 113.28 - 0.367014 T + 3.606 \cdot 10^{-4} T^2$ \cite{Gagliardi2007, hartel2015heat} (for which $\lambda_B$ hardly varies between 250 and \SI{350}{\kelvin}) but found only minor differences (\sfig{fig:nonpol:vareps}). 
We note that for narrow tubes, the relative permittivity may depend on the tube radius and differ from its bulk value (and also along and perpendicular to the tube axis). 
In the case of water, molecular dynamics (MD) simulations suggest substantial differences only for nanotubes smaller than about \SI{1}{\nano\meter} \cite{Schlaich2016, Loche2019, Loche2020}. 
Lacking a simple theory for relative permittivity in confinement, we followed \cit{hartel2015heat} and assumed that $\varepsilon_r$ is independent of the pore radius and equal to the bulk dielectric constant.

Besides through the electrostatic interactions (\cref{eq:Uel:approx}), ions mutually interact through hard sphere interactions 
\begin{align}
	\label{eq:HS}
	\beta \uu^{\rm HS}_{\alpha\gamma} (z)=
    \begin{cases}
		0, & z>2 \ionr\\
            \infty, & z\le 2\ionr\,.
	\end{cases}
\end{align}
The total ion-ion interaction energy now amounts to $\uu_{\alpha\gamma}=\uu^{\rm el}_{\alpha\gamma}+\uu^{\rm HS}_{\alpha\gamma}$.

\subsubsection{Ion-nanopore wall interactions}

We consider ions to interact with the pore wall through steric repulsions and attractive image-charge and dispersion forces---the latter were neglected in \cite{Verkholyak2021}.
The interaction energy due to image charges is \cite{rochester:cpc:13}
\begin{multline}
	\label{eq:Uself}
	\beta \uu_\image (r) = 
	 \frac{\lB}{2 \pi \poreri} \sum_{n=0}^\infty a_n \int_0^{2\pi} \dif \phi\, \cos(n\phi) \\
	\times \int_0^\infty \dif\xi\, \frac{I_n (\xi r/\poreri)}{I_n(\xi)} K_0 \left(\frac{\xi}{r}\sqrt{r^2 + \poreri^2-2r\poreri \cos\phi}\right),
\end{multline}
where $r$ and $\phi$ are radial and azimuthal coordinates, $a_n = 1$ if $n = 0$ and $a_n = 2$ otherwise, and $I_n(x)$ and $K_n(x)$ are the modified Bessel functions of the first and second kind, respectively. 

We calculated the dispersion (van der Waals) interaction energy between an ion and the nanotube wall by integrating atom-atom dispersion interactions (\ie, the long-range part of the Lennard-Jones potential) over the nanotube surface
\begin{equation}
\beta\uu_\vdW (r) =-4 \rhoc \epsLJ R\int_{-\infty}^{\infty}dz\int_0^{2\pi}d\phi\left(\frac{a+\carbonr}{D}\right)^{6},
\label{LJinten}
\end{equation}
where $\epsLJ$ is the Lennard-Jones parameter, $z$ is the axial coordinate, $\rhoc$ is the (2D) density of wall atoms (for a carbon nanotube $\rhoc=2/A_\mathrm{uc}=0.382$~\AA$^{-2}$, where $A_\mathrm{uc}=5.24$~\AA$^{2}$ is the area of the unit cell of a carbon monolayer, see \mycite{enoki2013book}), and $D=\left(R^2+r^2-2Rr\cos\left(\phi\right)+z^2\right)^{1/2}$ is the distance between an ion at position $r$ from the nanotube center and a point on the pore surface defined by $(R, \phi, z)$ in cylindrical coordinates (\cref{fig:model}d).
After integrating over $z$, we obtain
 \begin{align}
	\beta \uu_\vdW (r)
	= - \frac{3 \pi \porer \rhoc \epsLJ}{2} \int_0^{2\pi}\dif\phi\,
	 \frac{(\carbonr + \ionr)^{6}}{\left(\porer^2+r^2-2\porer r\cos\phi\right)^{5/2}}.
\label{eq:ULJ}
\end{align}

\label{ref:1:u}Note that neither $\uu_\image$ nor $\uu_\vdW$ depends on the ionic valency, so neither does the total ion-wall interaction, $\uu_{\pm}=\uu= \uu_\image + \uu_\vdW$. To account for the steric repulsions between the hard sphere ions and the pore wall, rather than adding an interaction energy term to $u$, we restrict the integration limits in the integral in \cref{eq:mu_shift} below.

\subsection{Mapping to a 1D model and its exact solution}
\label{sec:model:1D}

For narrow pores that can only accommodate a single file of ions, the above model can be mapped onto a one-dimensional model with effective electrochemical potentials given by \cite{Verkholyak2021}
\begin{align}
	\label{eq:mu1D}
	\muoneD_\pm = \mu_\pm - \bar \uu_\pm, 
\end{align}
where $\mu_\pm$ is the bulk chemical potential of cations ($+$) and anions ($-$), and 
\begin{align}
	\label{eq:mu_shift}
	\beta \bar \uu_\pm = 
	- \ln \left( 2 \pi \int_0^{\porera -\ionr} \e^{-\beta \uu_\pm(r)} r \,\dif r \right)
\end{align}
results from integrating over the degrees of freedom perpendicular to the nanopore symmetry axis \cite{Verkholyak2021}, where $\porera = \porer - \carbonr$ is the accessible pore radius.
\label{ref:3:HC}Note that the upper limit in the integral in \cref{eq:mu_shift} is due to hard-core exclusion between an ion and the pore wall.
Moreover, $\bar \uu_\pm$ has the unit of $\textrm{energy} \times \ln(\textrm{length}^2)$; since we included the thermal de Broglie wavelength in the chemical potential $\mu$, which thus has a term $\propto \textrm{energy}\times \ln(\textrm{length}^3)$, the resulting $\muoneD$ in \cref{eq:mu1D} contains a term proportional to energy $\times \ln (\textrm{length})$. 
Hence, $\exp(\muoneD/\kBT)$ has the unit of length.

Unlike the ion-ion interactions, which we assumed to be independent of the radial positions of the ions, in \cref{eq:mu_shift} we effectively took these degrees of freedom into account in the ion-pore wall interactions.
This turned out essential to reach a quantitative agreement with 3D Monte Carlo simulations \cite{Verkholyak2021}.

Given the exponential screening of ion-ion interactions by the electrons (and holes) on the nanotube walls (\cref{eq:Uel:approx}), interactions beyond nearest neighbours can be neglected. 
For any one-dimensional system of particles interacting solely with their nearest neighbours, one can analytically determine the partition function, particle densities, \etc \cite{longuet1958,Verkholyak2021}.
The calculations are lengthy and we do not present them here (see \cit{Verkholyak2021}). 
The result for the ion densities is (\ssect{sec:exact})
\begin{align}
\label{eq:rho:s}
\rho_\pm (s)
=-\frac{1}{2} \frac{\e^{\beta \muoneD}(\eta_{++}^2-\eta_{+-}^2) -\e^{\pm \beta e \Psi} \eta_{++}}
{\e^{\beta \muoneD}(\eta_{++}\eta'_{++}-\eta_{+-}\eta'_{+-}) - \cosh(\beta e \Psi)\eta'_{++}},
\end{align}
where $\kBT s$ is the 1D pressure, $\eta'_{\alpha\gamma} = \partial \eta_{\alpha\gamma}/\partial s$ and 
\begin{align}
\label{eq:eta}
\eta_{\alpha\gamma}(s) = \int_0^\infty \dif z\, \e^{-sz - \beta u_{\alpha\gamma}(z)}.
\end{align}
From this parametric solution, we can determine the surface charge density of the pore $\sigma = e(\rho_+ - \rho_-)/(2\pi\porer)$ at a given potential $\Psi$ and, in turn,  the differential capacitance $C(\Psi,T) = \partial \sigma/\partial\Psi$. 

\label{ref:3:1Dvs3D}\citeauthor{Verkholyak2021} \cite{Verkholyak2021} have compared the accumulated charge and capacitance obtained within this model with the results of 3D Monte Carlo simulations in a wide range of model parameters and found a quantitative agreement for pore sizes $\lesssim 1.4 \ionr$. 
For larger pores (but smaller than two ion radii), the 1D model still provided a qualitatively correct behaviour.

\subsection{Bulk chemical potential of ions}
\label{sec:model:MD}

The in-pore ion densities (\cref{eq:rho:s}) depend---besides on temperature, pore radius, applied potential difference, and Lennard-Jones parameter---on the bulk chemical potential of ions $\mu_\pm$ (see \cref{eq:mu1D}). 
To compute $\mu_\pm$, which also depends on temperature, we used the Espresso MD simulation package (\url{https://github.com/espressomd}) \cite{Weik2019} combined with the Widom insertion method \cite{widomMC:1963}. 
In all simulations, we set the ion size to $\ionr = \SI{3}{\angstrom}$ and the number of ions to $N_\pm = 100$. \label{ref:1:1} Moreover, we set ion concentration to $\SI{1}{M}$, which is typical for supercapacitors \cite{forse:jacs:16:chmec,hartel2015heat}; this concentration corresponds to the simulation box length $\approx \SI{55}{\angstrom}$. 
To mimic the hard-core interactions between the ions \cite{Jover2012}, we used the generic Lennard-Jones interaction potential with exponents $50$ and $49$ and the potential depth $\beta \epsilon_\mathrm{HS} = 1.5$.
As discussed below \cref{eq:Uel:approx}, we used a constant $\varepsilon_r=35$ in most calculations and checked a few results with a temperature-dependent relative permittivity $\varepsilon_r(T)$ taken the same as inside the pore.
The total number of iterations was $10^5$, with $10^3$ Widom insertions and $10^4$ MD steps per iteration. The results for the total chemical potential for a few temperatures are shown in \sfig{fig:mu}.

\section{Results and discussions}
\label{sec:results}
%
\subsection{Pore ionophilicity}
\label{sec:res:nonpol}

\begin{figure}
\includegraphics[width=\linewidth]{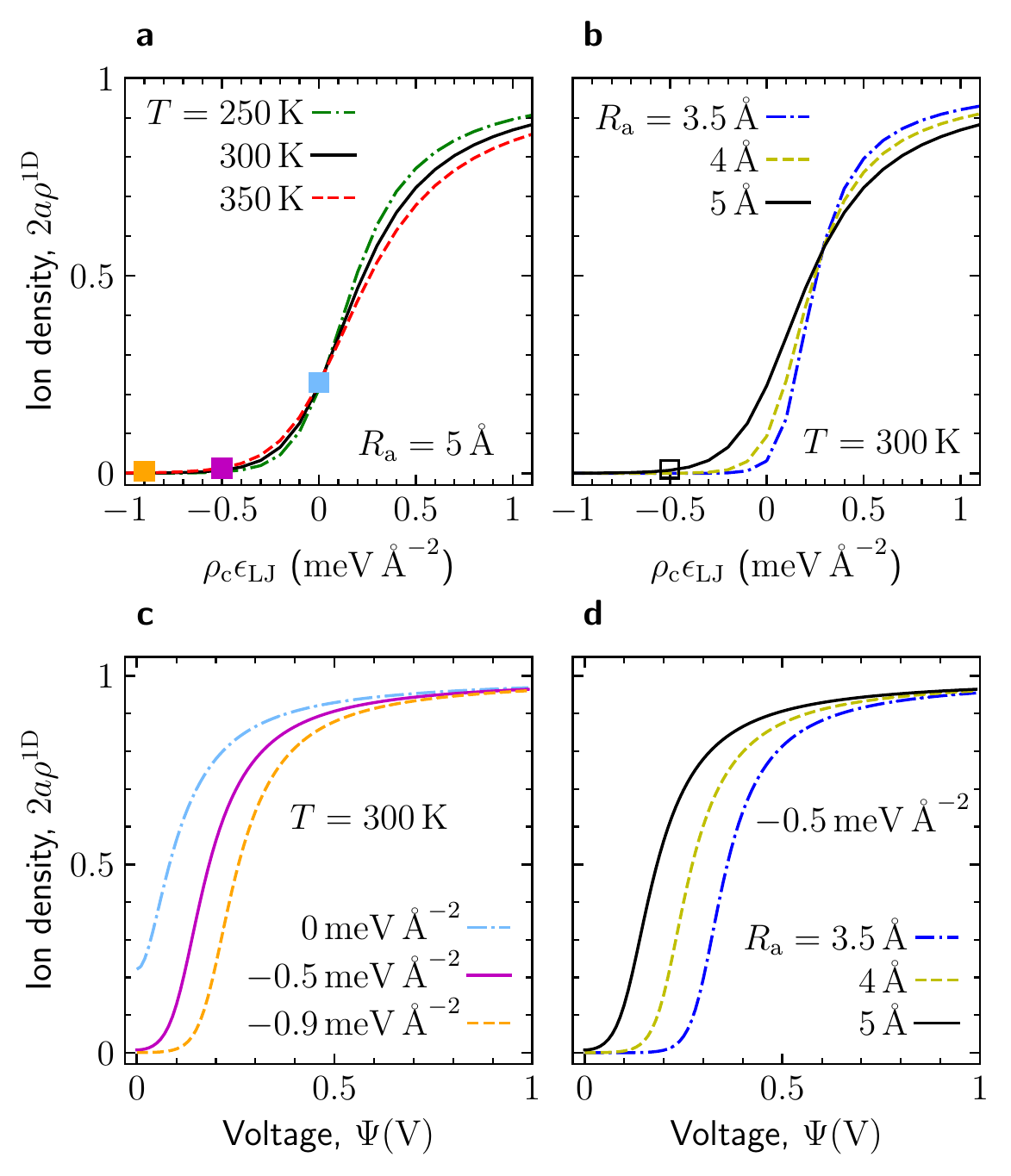}
\caption{\ftitle{Pore ionophilicity.} 
	\fsub{a,b} Total ion density inside a nonpolarized pore as a function of the Lennard-Jones parameter $\epsLJ$ for \fsub{a} accessible pore radius $\porera= \porer - \carbonr = \SI{5}{\angstrom}$ ($\carbonr = \SI{1.685}{\angstrom}$ is the radius of the nanotube wall atoms) and three temperatures and \fsub{b} for temperature $T=\SI{300}{\kelvin}$ and three pore radii. 
	The symbols in (a) indicate the values of $\epsLJ$ used in panel (c) and in \cref{fig:ionophil,fig:ionophil:voltage}. The symbol in (b) shows the value of $\epsLJ$ used in panel (d).
	\fsub{c,d} Total ion density inside a pore as a function applied potential difference $\Psi$ for \fsub{c} accessible pore radius $\porera= \SI{5}{\angstrom}$, temperature $T=\SI{300}{\kelvin}$ and three values of $\epsLJ$, and \fsub{d} for temperature $T=\SI{300}{\kelvin}$, $\rhoc\epsLJ=\SI{-0.5}{\milli \electronvolt\per\angstrom\squared}$ and three pore radii. 
	\label{fig:nonpol}
	}
\end{figure}

We first study pore ionophilicity, that is, the extent to which they are filled with ions when no potential is applied ($\Psi=\SI{0}{\volt}$). 
\Cref{fig:nonpol}a shows that the ion density in the pore increases from practically zero (ionophobic pores) to close to its maximal possible value, $\rhomax^\oneD = (2\ionr)^{-1}$, upon increasing the Lennard-Jones parameter $\epsLJ$, describing the pore-ion dispersion interactions. 
The data for the three widely-different temperatures considered here almost collapse; hence, the pore's ionophilicity depends sensitively on $\epsLJ$ but not on $T$.
\Cref{fig:nonpol}b shows that the pore radius also affects the pore's ionophilicity, which is due to both dispersion and image-charge interactions.
For $\epsLJ=0$, for instance, a \SI{3.5}{\angstrom} pore is ionophobic (vanishing ion density) while a \SI{5}{\angstrom} pore is filled to about a quarter of its maximal density.

In \cref{fig:nonpol}c, we plot the total ion density as a function of the applied potential difference $\Psi$ demonstrating that, as expected, ionophobic pores adsorb ions less readily than ionophilic pores.
This figure also shows that a larger potential must be applied to fill pores that are characterised by more negative ion-wall dispersion interaction parameter $\epsLJ$.
\Cref{fig:nonpol}d shows the voltage-induced pore filling of three pores with different radii but equal $\epsLJ$, as indicated with a square in panel (b) of this figure.
Although all three pores are ionophobic and practically free of ions at zero voltage, the narrowest pore requires a much higher voltage to start filling it.

\begin{figure}
\includegraphics[width=\linewidth]{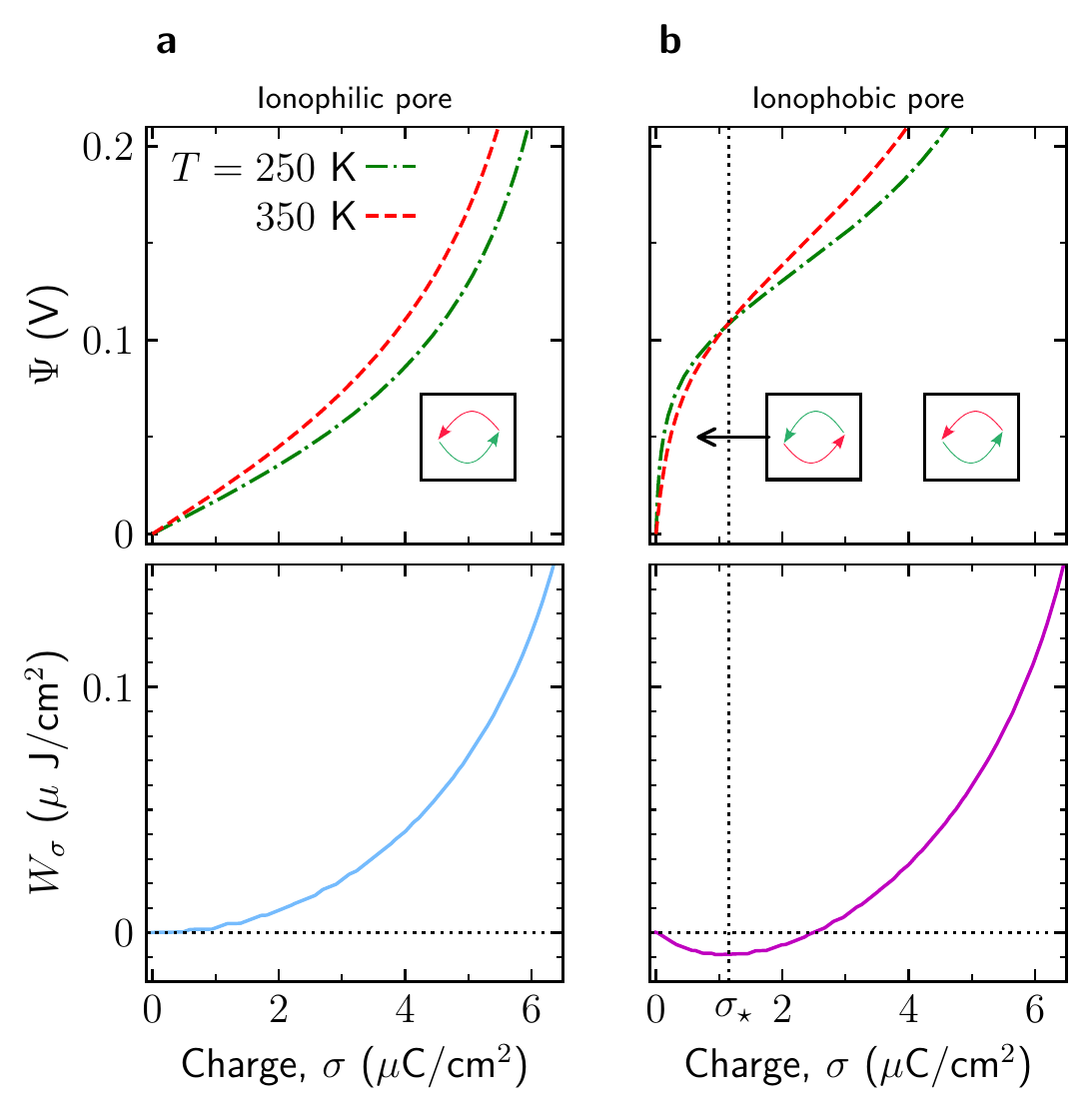}
\caption{\ftitle{HEC with ionophilic and ionophobic pores.} 
	\fsub{a}  Voltage for two temperatures (top) and harvested work during a Stirling-like charging cycle [\cref{eq:work:sigma}, using $\sigma_0=0$ and $\sigma_1=\sigma$] (bottom) \vs the surface charge density. 
	We set the accessible pore radius to $\porera=\SI{5}{\angstrom}$, the ion radius to $\ionr=\SI{3}{\angstrom}$, and the Lennard-Jones parameter to $\epsLJ=0$---corresponding to an ionophilic pore (\cref{fig:nonpol}a).
	\fsub{b} The same as (a) but for $\rhoc \epsLJ=\SI{-0.5}{\milli \electronvolt\per\angstrom\squared}$, for which the pore is ionophobic (\cref{fig:nonpol}a). 
	The pore has an anomalous Seebeck coefficient $\mathcal{S}<0$ at small surface charge densities. 
	We replot this figure in \cref{sm:fig:ionophil} in the \ESIname for a pore for which $\porera=\SI{4}{\angstrom}$. 
	\label{fig:ionophil}
	}
\end{figure}

\subsection{HEC with ionophilic and ionophobic pores}

To study how a pore's ionophilicity affects HEC, we consider a pore with an accessible radius $\porera=\SI{5}{\angstrom}$ and two values of the Lennard-Jones parameter, $\epsLJ = 0$ and $\rhoc \epsLJ = \SI{-0.5}{\milli \eV\per\angstrom\squared}$, corresponding to a moderately ionophilic and ionophobic pore, respectively (symbols in \cref{fig:nonpol}a). 
Note that for these $\epsLJ$ values, the ion density hardly changes with temperature (\cref{fig:nonpol}a), \ie, the pores remain ionophilic or ionophobic as we change the temperature.

The top panel of \cref{fig:ionophil}a shows charging curves of the ionophilic pore. 
We see that, for all surface charge densities $\sigma$ considered, the electrode potential $\Psi$ increases with temperature (\ie, $\mathcal{S}>0$).
In this case of a positive ionic Seebeck coefficient, a charging-heating-discharging-cooling cycle yields positive work.
The bottom panel shows that the work output $W_\sigma$ of a Stirling-like cycle (\cref{eq:work:sigma}, using $\sigma_0=0$ and $\sigma_1=\sigma$) grows monotonously with $\sigma$.

The ionopobic pore ($\rhoc \epsLJ =\SI{-0.5}{\milli\eV\per\angstrom\squared}$) exhibits a negative Seebeck coefficient at small $\sigma$ and a positive Seebeck coefficient at large $\sigma$ (\cref{fig:ionophil}b top panel), with a crossover around $\sigma_{\star} \approx \SI{1.1}{\micro\coulomb\per\centi\meter\squared}$. 
In the bottom panel of \cref{fig:ionophil}b, we again plot the work output $W_\sigma$ of a Stirling-like charging-heating-discharging-cooling cycle (\cref{eq:work:sigma}, using $\sigma_0=0$ and $\sigma_1=\sigma$).
The work output is now non-monotonous, with a negative minimum at $\sigma_{\star}$ and a monotonous increase beyond $\sigma_{\star}$. 
Hence, to harvest energy at small surface charge densities ($\sigma<\sigma_{\star}$), one should perform a reversed charging-cooling-discharging-heating cycle instead.
Alternatively, a regular charging-heating-discharging-cooling cycle starting at $\sigma_0=\sigma_{\star}$ harvest energy for any $\sigma_1>\sigma_0$ as well.

\subsection{Nanopore ionophilicity and Seebeck coefficient}

\Cref{fig:ionophil} revealed that the Seebeck coefficient $\mathcal{S}$ flips sign for small $\sigma$ upon changing $\epsLJ$ [\cf panels (a) and (b)] and, for small $\epsLJ$, upon changing $\sigma$ [panel (b)].
To understand these sign flips, we present analytical expressions for the electrostatic potentials for opposite limits of the surface charge densities $\sigma$.
For $\sigma$ close to its maximum value, $\sigma_\mathrm{max} = e/(4\pi \porer\ionr)$, we have \cite{Verkholyak2021}
\begin{align}
	\label{eq:sigma_high}
	\Psi(\sigma,T)\approx\frac{1}{2\ionr\beta}\frac{1}{\sigma_\mathrm{max}-\sigma}.
\end{align}
Clearly, \cref{eq:sigma_high} predicts $\mathcal{S}>0$ for all temperatures, surface charge densities, and independent of the pore ionophilicity. 
For small $\sigma$, we find in \ssect{sec:low-density} that
\begin{align}
\label{eq:C0}
	\Psi(\sigma,T) \approx \sigma\frac{\pi \porer }{\beta e^2}\left[8 \ionr + \exp\left(-\beta\muoneD\right)\right].
\end{align}
(Note that $\exp(-\beta\muoneD)$ has the unit of length, see below \cref{eq:mu_shift}.)
For point ions ($\ionr =0$), which is a reasonable simplification at low densities, it is not difficult to see that 
$\mathcal{S} < 0$  for $\muoneD < - \kBT$ and $\mathcal{S} > 0$ otherwise.
Numerical experimentation with \cref{eq:C0} confirms that indeed $\mathcal{S}$ is negative (positive) for ionophobic (ionophilic) pores.
Thus, crossovers from $\mathcal{S}<0$ to $\mathcal{S}>0$ happen for ionophobic pores at an intermediate $\sigma$ and for pores with small $\sigma$ at an intermediate $\epsLJ$ value.

While we derived \cref{eq:C0} from an exact analytical expression for ion densities, \cref{eq:rho:s} (see \ssect{sec:low-density}), it can also be obtained using a model of charged particles in 1D interacting solely through hard-core repulsion; the free energy density of this model is given by (\ssect{sec:noninter_gas})
\begin{align}
	\label{eq:f}
	f(\rho_\pm, \Psi) = \sum_\alpha \rho_\alpha (\mu_\alpha \pm e \Psi ) - T s (\rho),
\end{align}
where $s(\rho)$ is the entropy density, which depends on the total ion density $\rho=\rho_+ + \rho_-$ and is known exactly in 1D \cite{Tonks1936Complete}. In \cref{eq:f}, we neglected the ion-ion interactions because they are exponentially screened by the nanotube wall (see \cref{sec:model:Uij}). \Cref{eq:C0} follows from the minimum of $f$ determined by $\partial f / \partial\rho_\alpha = 0$, assuming low ion densities and small $\Psi$.

Interestingly, an equation similar to \cref{eq:C0} with similar properties follows from \cref{eq:f} also for narrow slit pores, if we use for $s(\rho)$ the scaled particle results for a 2D hard-disk system \cite{Helfand1961Theory, Holovko2010Analytical}  (\ssect{sec:noninter_gas:2D}). 
Note that for point ions, $s$ is given by the ideal-gas entropy for both 1D and 2D pores, leading to the same behaviour. In a more general case, a similar numerical experimentation as for \cref{eq:C0} shows that $\mathcal{S} < 0$ for ionophobic and $\mathcal{S} >0$ for ionophilic 2D pores (\ssect{sec:noninter_gas:2D}). 
Since an equation similar to \cref{eq:sigma_high} can also be developed for slit pores \cite{Verkholyak2021}, giving a positive Seebeck coefficient independently of the model parameters, there must also be a crossover between $\mathcal{S}<0$ and $\mathcal{S}>0$ for ionophobic pores induced by varying the surface charge.

These results, therefore, indicate that the crossovers between negative and positive Seebeck coefficients might be a generic feature of narrow pores, independent of their geometry.

\begin{figure}[t]
\includegraphics[width=\linewidth]{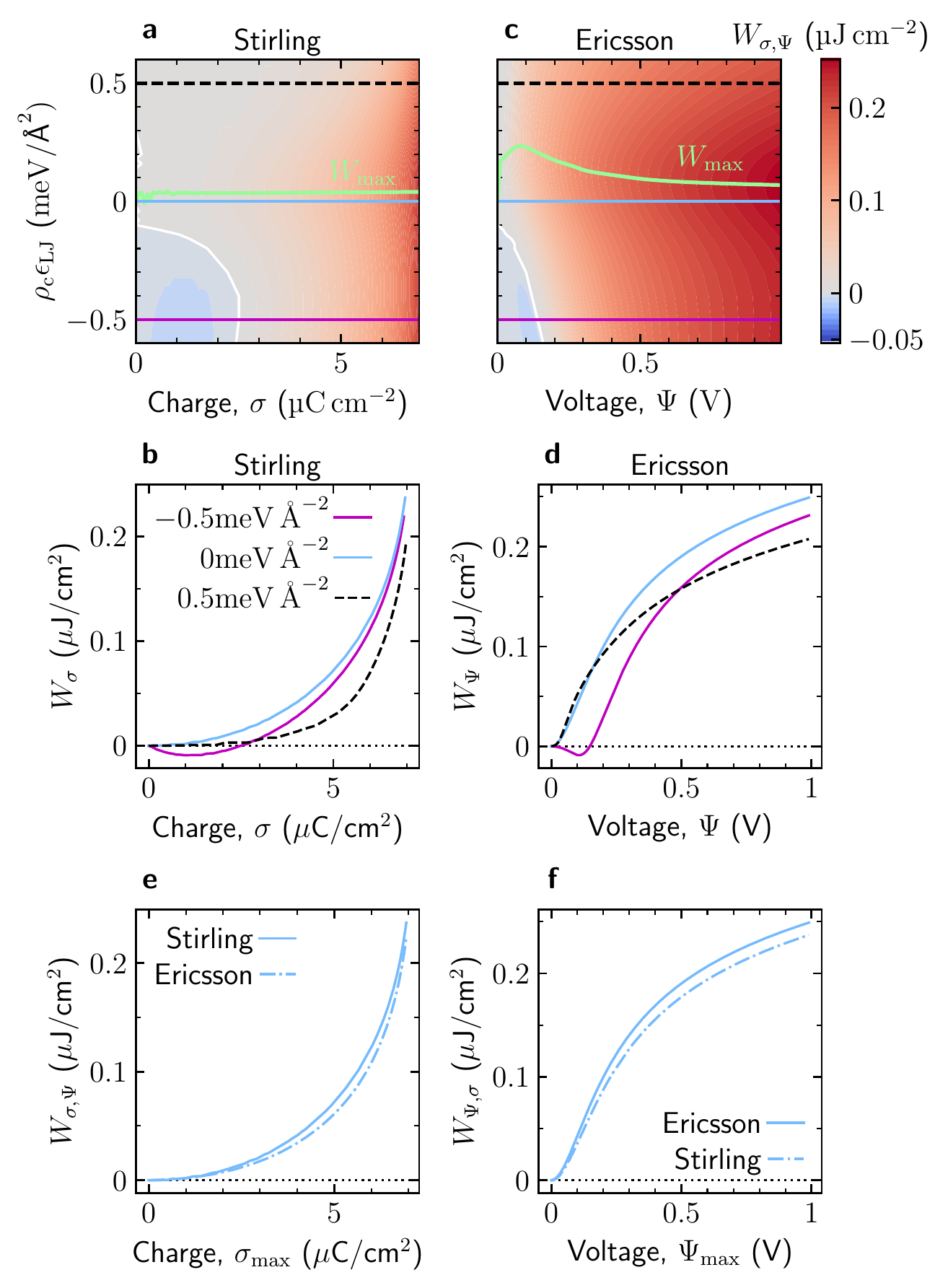}
\caption{\ftitle{Energy harvesting with Stirling and Ericsson-like cycles.}
	\fsub{a} Heatmap showing the energy harvested with Stirling-like cycles in the plane of $\epsLJ$ and $\sigma$. The horizontal lines show the values of $\epsLJ$ used in \cref{fig:ionophil} and in \fsub{b}. The green line shows $\epsLJ$ providing the maximum harvested energy $W_\mathrm{max}$ as a function of $\sigma$.
	\fsub{c} Heatmap showing the energy harvested with Ericsson-like cycles in the plane of $\epsLJ$ and $\Psi$. 
	The horizontal lines show the values of $\epsLJ$ used in \fsub{d}. The green line shows $\epsLJ$ providing the maximum energy $W_\mathrm{max}$ as a function of $\Psi$.
	\fsub{e,f} Comparison of Stirling and Ericsson-like cycles in the case when there is a limitation on \fsub {e} maximum accumulated charge and \fsub{f} maximum applied potential difference. 
	The ion radius $\ionr=\SI{3}{\angstrom}$ and the accessible pore radius $\porera=\SI{5}{\angstrom}$ in all panels. 
	For a narrower pore, see \cref{sm:fig:ionophil,sm:fig:ionophil:voltage} in the \ESIname.
	\label{fig:ionophil:voltage}
	}
\end{figure}

\subsection{Dependence on ion-wall dispersion interactions}

To scrutinize the ramifications of these two $\mathcal{S}$ crossovers, we plot in \cref{fig:ionophil:voltage}a the work $W_\sigma$ done during a Stirling-like cycle (\cref{eq:work:sigma}, using $\sigma_0=0$ and $\sigma_1=\sigma$) in the plane of $\epsLJ$ and $\sigma$. This plot confirms that ionophobic pores (large negative $\epsLJ$) have negative $W_\sigma$ at small $\sigma$, requiring a reversed charging-cooling-discharging-heating cycle to harvest net positive energy. 

In \cref{fig:ionophil:voltage}b, we show three cuts through the heatmap of panel (a) at constant $\epsLJ$.
As discussed, the ionophobic pore ($\rhoc\epsLJ=\SI{-0.5}{\milli \electronvolt\per\angstrom\squared}$) loses energy during charging cycles at small $\sigma$. 
As $\sigma$ increases, however, it provides comparable and even larger harvested energies than the ionophilic pores. 
This figure also demonstrates that strongly ionophilic nanopores can make energy harvesting less effective.
Indeed, the $\epsLJ$ value that maximises $W_\sigma$ depends on $\sigma$ but always lies close to zero (the green line in \cref{fig:ionophil:voltage}a). 
By ignoring wall-ion dispersion interactions, \citeauthor{hartel2015heat}\cite{hartel2015heat} thus fortuitously studied a close-to-optimal parameter setting.

\subsection{Stirling \vs Ericsson-like cycles}

In \cref{fig:ionophil:voltage}c, we plot the work $W_\Psi$ done during an Ericsson-like cycle (\cref{eq:work:sigma}, using $\Psi_0=0$ and $\Psi_1=\Psi$) in the plane of $\epsLJ$ and $\Psi$.
In panel (d), we again show a cut through the heatmap in (c) for three values of $\epsLJ$. 
The work output of ionophilic and ionophobic pores differs more for Ericsson-like than for Stirling-like cycles.
At high $\Psi \gtrsim \SI{0.2}{\volt}$, the pore with $\epsLJ=0$ clearly outperforms both ionophobic and strongly ionophilic pores.
The green line in \cref{fig:ionophil:voltage}c shows that, at high $\Psi$, the value of $\epsLJ$ maximising the harvested energy is indeed close to zero.
Comparing \cref{fig:ionophil:voltage}a and c, we see that Stirling and Ericsson-like cycles harvest the maximal energy for slightly different values of $\epsLJ$.
In both cases, however, the optimal green lines lie close to $\epsLJ = 0$, which, for the considered pore radius $\porera=\SI{5}{\angstrom}$, corresponds to moderate ionophilicity (ion density at zero voltage $2\ionr\rho^\oneD \approx 0.3$, see \cref{fig:nonpol}b).
We found a similar behaviour also for narrower pores (\sfig{fig:ionophil}).

To directly compare Stirling and Ericsson-like cycles, we plot energies harvested with both cycles when there are limitations on the accumulated charge ($\sigma < \sigma_\mathrm{max}$) and potential difference ($\Psi < \Psi_\mathrm{max}$) of a supercapacitor (\cref{fig:ionophil:voltage}e and f, respectively). 
As anticipated in \cref{sec:general}, Stirling and Ericsson-like cycles give higher work outputs when operating under charge and voltage limitations, respectively.
However, when the charge is bound by $\sigma_\mathrm{max}$, the difference between the two cycles practically vanishes at large $\sigma_\mathrm{max}$. 
In contrast, this difference increases with $\Psi_\mathrm{max}$ when working under voltage limitations.

\subsection{Effect of pore size}

\begin{figure}
\includegraphics[width=\linewidth]{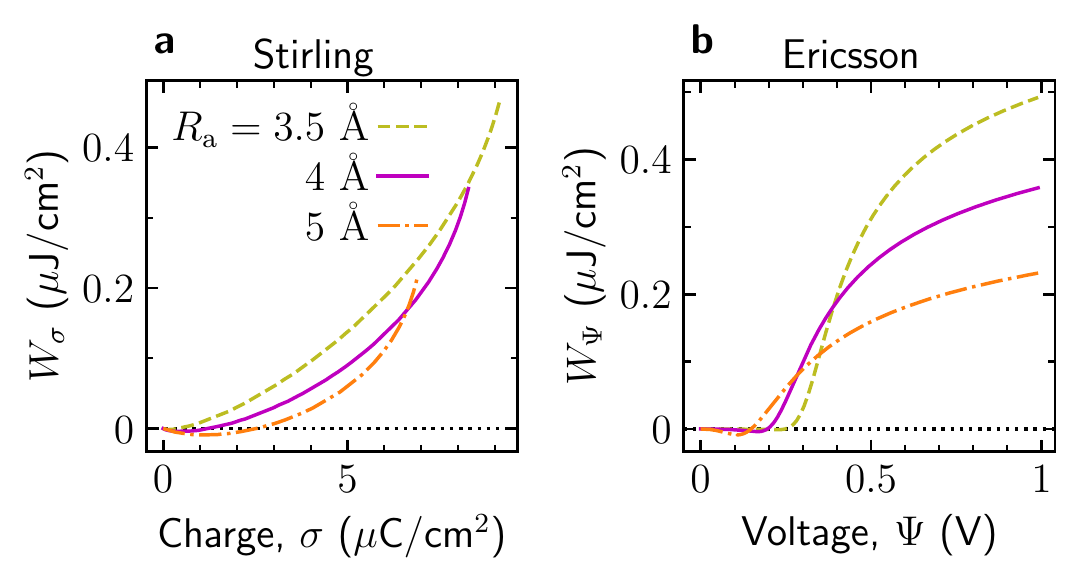}
\caption{\ftitle{Effect of pore size on HEC by ionophobic pores.} 
	Harvested energy as a function of \fsub{a} charge and \fsub{b} voltage for a few pore sizes. We set the ion radius to $\ionr=\SI{3}{\angstrom}$ and the Lennard-Jones parameter to $\rhoc \epsLJ=\SI{-0.5}{\milli\electronvolt\per\angstrom\squared}$, corresponding to an ionophobic pore. 
    For examples of the pore-size dependence of energy harvesting for larger $\rhoc \epsLJ$, giving moderately and strongly ionophilic pores, see \cref{sm:fig:pore_size:moderate,sm:fig:pore_size:philic} of \ESIname.
	\label{fig:pore_size:phobic}
	}
\end{figure}

\Cref{fig:pore_size:phobic} shows the harvested energy of Stirling and Ericsson-like cycles for pores with $\rhoc \epsLJ=\SI{-0.5}{\milli\electronvolt\per\angstrom\squared}$ and three radii $\porer$ (for two larger $\rhoc \epsLJ$ values, see  \cref{sm:fig:pore_size:moderate,sm:fig:pore_size:philic}).
All three pores are ionophobic, though the narrowest pore more so than the wider ones (\cref{fig:nonpol}d).

For the Stirling-like cycle, the region of anomalous behaviour shrinks with decreasing pore radius, though a small $W_\sigma<0$ region remains even for the narrowest pore considered ($\SI{3.5}{\angstrom}$).
For the Ericsson-like cycle (\cref{fig:pore_size:phobic}b), the region of negative $W_\Psi$ extends to higher voltages, while the minimum in $W_\Psi$ decreases.
For instance, the harvested energy for the smallest (\SI{3.5}{\angstrom}) pore is roughly zero until about $\SI{0.3}{\volt}$ and then starts increasing rapidly. 
This behaviour is because strongly ionophobic pores subject to small applied potentials are hardly charged and show only a weak dependence of the accumulated charge on temperature, leading to low harvested energies until a sufficiently high voltage is applied for ions to overcome the barrier imposed by ionophobicity and enter the pore (\cref{fig:nonpol}c,d).

The smallest pore being most ionophobic among the three pores of \cref{fig:pore_size:phobic} and harvesting most energy at large potential differences suggests a positive correlation between ionophobicity and energy harvesting.
However, when we varied the pore’s ionophilicity by varying $\epsLJ$ at a fixed pore size, the moderately ionophilic pore performed as good or better than ionophilic and ionophobic pores for all $\Psi$ (\cref{fig:ionophil:voltage}a-d).
Thus, although the narrowest and hence most ionophobic pore yielded the most harvested energy for $\rhoc \epsLJ=\SI{-0.5}{\milli\electronvolt\per\angstrom\squared}$ as used in \cref{fig:pore_size:phobic}, moderately ionophilic pores with $\epsLJ$ close to zero can harvest even more work (\sfig{fig:epsLJ}).

\section{Conclusions}
\label{sec:concl}

Classical heat engines convert heat into mechanical work by cyclically expanding and contracting at different temperatures.
In the same way, a supercapacitor can convert heat into electricity by cyclically charging and discharging at different temperatures. 
The order in which the supercapacitor should charge, discharge, heat, and cool depends on its Seebeck coefficient, $\mathcal{S}(\sigma,T)=\partial \Psi/\partial T$, with $\Psi(\sigma,T)$ the pore's potential when it carries a surface charge density $\sigma$.
Herein, we determined $\Psi(\sigma,T)$ through an analytically-solvable single-file cylindrical pore model. 
The extent to which the pore was filled when no potential was applied, its \textit{ionophilicity}, depended mostly on the pore's radius and the dispersion interactions between the pore and the ions (\cref{fig:nonpol}). 
We found $\mathcal{S}>0$ for ionophilic pores, in line with experiments on commercial supercapacitors.
Conversely, ionophobic pores yielded $\mathcal{S}<0$, but only for small surface charge densities $\sigma$, with a crossover to the $\mathcal{S}>0$ behaviour as $\sigma$ increased (\cref{fig:ionophil}). 
We provided arguments that similar crossovers also occur for slit nanopores, suggesting that this behaviour is a generic feature of narrow pores that does not depend on the pore geometry. 
Thus, the Seebeck coefficient may serve as an indicator of a pore's ionophobicity.

In analogy to the Stirling and Ericsson cycles of classical heat engines, we considered two types of cycles wherein the pore was heated and cooled either at a constant surface charge density or constant potential. 
Both types of cycles yielded maximal work output for narrow pores and weak dispersion interactions, corresponding to moderately ionophilic pores (\cref{fig:ionophil:voltage,fig:pore_size:phobic}).
However, the Stirling and Ericsson-like cycles reached maximal work output for slightly different values of the ion-wall dispersion interaction strength (\cref{fig:ionophil:voltage}a,c).
We found that an Ericsson-like cycle is optimal when there are limitations on the applied potential, as is the case for supercapacitors.
When the operation is limited by accumulated charges, a Stirling-like cycle allows one to harness higher energies (\cref{fig:ionophil:voltage}e,f).

It is interesting to compare the parameters optimising a supercapacitor for heat-to-energy conversion with the parameters optimising it for energy storage. We found that narrow pores provide the highest achievable harvested energy (\cref{fig:pore_size:phobic}), similarly as they do for the capacitance and stored energy \cite{gogotsi:sci:06,pinero:carbon:06,kondrat:ees:12}. However, our calculations showed that moderately ionophilic pores harness the maximal energy, while ionophobic pores are not optimal for this purpose.
This behaviour contrasts with energy storage, which is maximised by ionophobic pores when operating at elevated voltages \cite{kondrat:nh:16,lian_wu:jpcm:16:Ionophobic}

Future work could account for the pore-network structure and pore-size distribution of supercapacitors, and could study out-of-equilibrium charging.
We hope our results motivate further simulation, experimental and engineering work to develop supercapacitor-based devices for the ecologically-friendly conversion of waste heat to electrical energy.

\begin{acknowledgements}
This work was supported by NCN grant No. 2021/40/Q/ST4/00160 to SK. 
\end{acknowledgements}

\bibliography{supercaps,exact,graphene,references,bibliography_MJ}

\end{document}